# Conetronics in 2D Metal-Organic Frameworks: Double Dirac Cones, Magnetic Half Dirac Cones and Quantum Anomalous Hall Effect

Menghao Wu[1,2*], Zhijun Wang[1], Junwei Liu[3], Huahua Fu[1], Lei Sun[4], Xin Liu[1], Minghu Pan[1], Hongming Weng[5], Mircea Dincă[4], Liang Fu[3], Ju Li[2*]

[1]School of Physics and Wuhan National High Magnetic Field Center, Huazhong University of Science and Technology, Wuhan, Hubei 430074, China

[2]Department of Nuclear Science and Engineering, Massachusetts Institute of Technology, Cambridge, MA 02139, USA

[3]Department of Physics, Massachusetts Institute of Technology, Cambridge, MA 02139, USA

[4]Department of Chemistry, Massachusetts Institute of Technology, Cambridge, MA 02139, USA

[5]Beijing National Laboratory for Condensed Matter Physics, and Institute of Physics, Chinese Academy of Sciences, Beijing 100190, China

Abstract: Based on recently synthesized $Ni_3C_{12}S_{12}$–class 2D metal-organic frameworks (MOFs), we predict electronic properties of $M_3C_{12}S_{12}$ and $M_3C_{12}O_{12}$, where M=Zn, Cd, Hg, Be, or Mg with no M orbital contributions to bands near Fermi level. For $M_3C_{12}S_{12}$, their band structures exhibit double Dirac cones with different Fermi velocities that are n (electron) and p (hole) type, respectively, which are switchable by few-percent strain. The crossing of two cones are symmetry-protected to be non-hybridizing, leading to two independent channels in 2D node-line semimetals at the same **k**-point akin to spin-channels in spintronics, rendering "conetronics" device possible. The node line rings right at their crossing, which are both electron and hole pockets at the Fermi level, can give rise to magnetoresistance that will not saturate when the magnetic field is infinitely large, due to perfect n-p compensation. For $M_3C_{12}O_{12}$, together with conjugated metal-tricatecholate polymers $M_3(HHTP)_2$, the spin-polarized slow Dirac cone center is pinned precisely at the Fermi level, making the systems conducting in only one spin/cone channel. Quantum anomalous Hall effect can arise in MOFs with non-negligible spin-orbit coupling like $Cu_3C_{12}O_{12}$. Compounds of $M_3C_{12}S_{12}$ and $M_3C_{12}O_{12}$ with different M, can be used to build spintronic and cone-selecting heterostructure devices, tunable by strain or electrostatic gating.

**Introduction**

It is well known that graphene possesses a band structure with Dirac cones touching at the Fermi level at each K point in the Brillouin zone, giving rise to massless Dirac fermions with rich physics [1]. Over the last few years, linear band dispersion and the associated Dirac physics have been explored in other materials[2], such as graphyne[3, 4] and the surface of 3D topological insulators[5]. Similar phenomena have also been investigated in artificial graphene, such as metal surfaces with hexagonal assemblies of CO molecules[6], or ultra-cold atoms trapped in honeycomb optical lattices[7]. For 2D or quasi-2D crystals with three-fold symmetry, generally, K and K' points of Brillouin zone are symmetry related, leading to two-fold degenerate bands generally with a Dirac-like dispersion at K and K'.[2] Recently, some designed 2D conjugated polymer[8] with three-fold symmetry have been predicted to have graphene-like electronic structures. Some structures with spin-splitting were shown to exhibit half-metallicity[9] or even quantum anomalous Hall effect (QAHE)[10-12] when spin-orbit coupling (SOC) is taken into account. Actually, $Ni_3C_{12}S_{12}$, a conjugated 2D nanosheet comprising of planar nickel bis(dithiolene) complexes that has been successfully synthesized by Kambe *et al*. [13, 14], also has three-fold symmetry and has shown high electrical conductivity (up to 160 S/cm at 300K) and controllable oxidation states (reduced using tetracyanoquinodimethane sodium (NaTCNQ) and oxidized using tris(4-bromophenyl)aminium hexachloroantimonate) [10], analogous to 2D graphene / graphene oxide. Band structure calculation shows that native undoped $Ni_3C_{12}S_{12}$ single-sheet is a semiconductor, while a topological insulator (TI) state is predicted within a bandgap of Dirac band opened up by SOC at around 0.5eV above the undoped Fermi level[15].

Herein, we report a series of 2D MOF semimetals in their native, undoped state with intriguing electronic and magnetic properties. Some of these proposed materials possess two independent Dirac cones of different Fermi velocity at each K point that are N (electron) and P (hole) type, respectively, and the relative energy alignment of fast and slow Dirac cones can be tuned via few-percent tensile strain so as to switch between N and P designation; for some others there is only one spin-polarized Dirac cone (half Dirac cone) crossing the Fermi level at each K point, making the systems ferromagnetic and conducting in only one spin channel and one cone channel. The latter are predicted to exhibit the QAHE when SOC is taken into account. Finally, we show that "conetronics" devices can be designed for manipulating carriers in different cones. Previous experimental reports have already shown that in $Ni_3C_{12}S_{12}$, Ni atoms may be replaced by other transition metal atoms while S atoms can be replaced by O or -NH groups[16-18], and some of them were also predicted to exhibit TI state around 0.5eV away from Fermi level[19, 20]. Here we mainly focus on $M_3C_{12}S_{12}$ and $M_3C_{12}O_{12}$ where M=Zn, Cd, Hg, Be, Mg. These metals strongly favor the $2^+$ formal oxidation state, and thus are expected to donate two s electrons to nearby four S (or O) atoms. They contribute no d electrons to the density of states near Fermi level and require no external chemical doping. We note that although square planar complexes of these metals are uncommon, their closed-shell electronic configurations do not favor any particular crystal field geometry, their coordination geometry being influenced either by steric effects or other global stabilization effects, such as strain or Coulombic interactions.

**Double Dirac cones of $M_3C_{12}S_{12}$**

Figures 1a-d display the band structures of $M_3C_{12}S_{12}$, where M = Zn, Cd, Hg, Be, Mg. They all exhibit two Dirac cones near the Fermi level at each K point, with different cone-tip energies that are separated

by the Fermi-level and with different Fermi velocity. For M = Zn, Cd, Hg, at each K point, the fast Dirac cone is found above the Fermi level (N-type) where the Dirac fermion carriers are excess electrons, while the slow Dirac cone is below the Fermi level (P-type) where the Dirac fermion carriers are holes, at the native undoped state. In contrast, for M = Be, Mg, the fast Dirac cone is P-type while the slow one is N-type. For the fast Dirac cones, $Hg_3C_{12}S_{12}$ has the highest Fermi velocity $V_F$ (approximately half of that of graphene) which is almost three times larger than that of $Mg_3C_{12}S_{12}$ (Supplementary Table S1), and four times larger than $V_F'$ of the slow Dirac cone. The band structures in Fig.1 exhibit closed band crossing lines by intersections of two bands, namely, node-line rings, right at the Fermi level, making those systems perfect 2D node-line semimetals, as shown on the right panel where those rings are marked by purple: for M=Zn, Cd, Be, there is a node-line ring locate around each K point due to the intersections of fast and slow Dirac cones, and in the displayed band structures we can find an intersection point of the fast and slow Dirac cone along the path Γ - K and K - M respectively. For the band structures of M = Hg, Mg, a closed ring around the $C_6$ rotation axis at Fermi level is formed by the band crossings, exhibiting linear dispersions along Γ–K and Γ–M. For all those systems, no gap opened at the crossing of two cones, indicating no hybridization, so the slow and fast Dirac cones can be regarded as two independent channels just like spin-up and spin-down in spintronics, which will be discussed for constructing "conetronics" devices in the last section. We have also checked the energy of buckled states which may exist in such 2D MOFs [21], and found that for M=Mg, its buckled state is quite close to the planar state in energy: upon a biaxial strain of -1.3%, the buckled state will be the ground state with a magnetic moment of $2\mu_B$ per supercell. The spin-splitting makes the system metallic only in the slow Dirac cone and in only one spin channel; upon a biaxial tensile strain of 1.3%, the system is planar but also becomes ferromagnetic. The band structures of these buckled systems are displayed in Supplementary Figure S1. If M is substituted by other alkali earth metal such as Ca, Sr, Ba, the buckled states are the ground states where S atoms will be more than 0.4 Å away from the MOF plane, which may indicate instability as their planar tetracoordinate (square-planar) structures are not favorable in energy. Here we only focus on M = Zn, Cd, Hg, Be, Mg where the optimized structures may reveal that their square-planar coordination are favorable in 2D periodic structures.

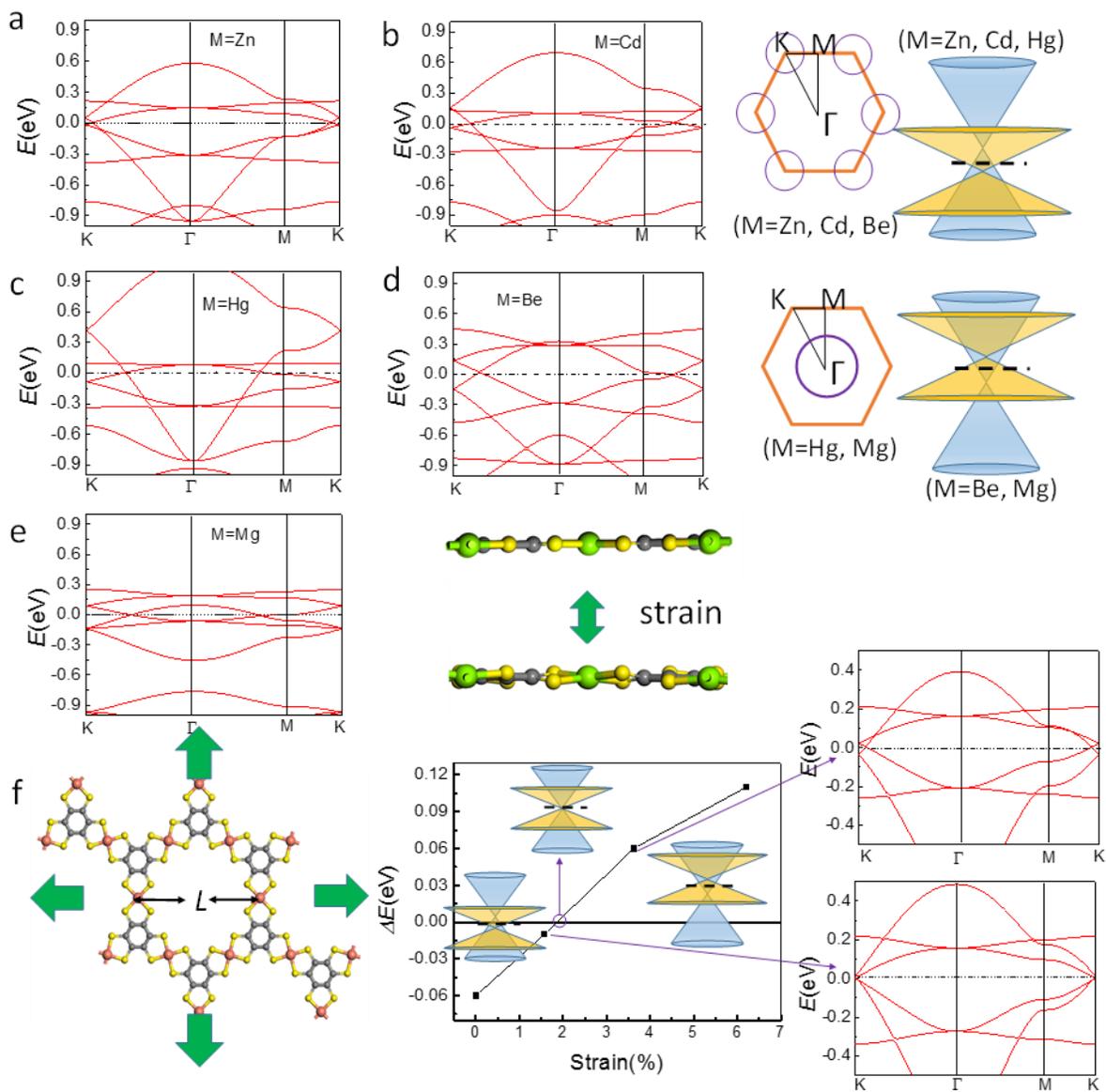

**Figure 1 Band structures of $M_3C_{12}S_{12}$**

(a-e) Band structures of $M_3C_{12}S_{12}$ for M = Zn, Cd, Hg, Be, and Mg, and their node-line rings in Brillouin zone. (f) Dependence of energy difference between the slow Dirac cone and fast Dirac cone on biaxial strain for $Zn_3C_{12}S_{12}$. The grey, yellow, green and red spheres denote carbon, sulfur, magnesium and copper atoms respectively.

Those systems are unique not only for their two Dirac cones with different velocities near the Fermi level. Each node ring right at the Fermi level is the electron pocket of one Dirac cone and meanwhile the hole pocket of the other cone leading to a perfect n-p compensation, which is hitherto unreported to our knowledge. The compensated electron and hole pockets can give rise to excitonic electron-hole pair condensation and high-temperature superconductivity[22]. Moreover, for a semimetal in a strong magnetic field[23], the magnetoresistance

$$\rho_{xx} = \frac{n^2\rho_e + p^2\rho_h}{(n^2\rho_e + p^2\rho_h)^2 \frac{(ec)^2}{B^2} + (n-p)^2},$$

where *n*, *p* are electron/hole concentration and $\rho_e, \rho_h$ are resistivity predicted by Drude theory. In a strong *B* field, if $n \neq p$, $\rho_{xx} \to \frac{n^2\rho_e + p^2\rho_h}{(n-p)^2}$, the magnetoresistance saturates to a field-independent value; if *n=p*, the magnetoresistance will increase as $B^2$ without saturating. Recently a breakthrough has been reported that $WTe_2$ possesses a large magnetoresistance that will not saturate even up to 60T, as it is the first known material with electron and hole pockets making up a nearly perfect n-p compensation[24]. Here in our systems, the electron and hole pockets correspond to the same node line rings, so the *n-p* compensation will be exactly perfect, and theoretically their magnetoresistance will not saturate even when *B* is infinitely large!

Since the orbitals of fast and slow Dirac cones refer to different resonance structures that favor different bond length patterns, it is possible to use strain to control the radius of rings formed by the crossings of fast and slow Dirac cones. For M = Zn and Cd, the fast and slow Dirac cones are both close to the Fermi level, giving rise the possibility to tune the relative energy alignment between fast and slow Dirac cones. Take M = Zn as an example, as shown in Fig. 1e, we define *ΔE* = *E*(slow) - *E*(fast) as the energy difference between the slow Dirac cone and fast Dirac cone tips. Upon a biaxial strain, the fast cone will shift down while the slow cone will move upwards. As the strain reaches around 1.9%, where *ΔE* = 0, two Dirac points coalesce exactly at the Fermi level, forming a Dirac point with two different Fermi velocities: with larger strain, the fast cone will switch to P-type while the slow one will become N-type. We note that 1.9% tensile strain should be achievable experimentally in most 2D monolayer materials without fracture [25].

**Figure 2 Electron states distribution**

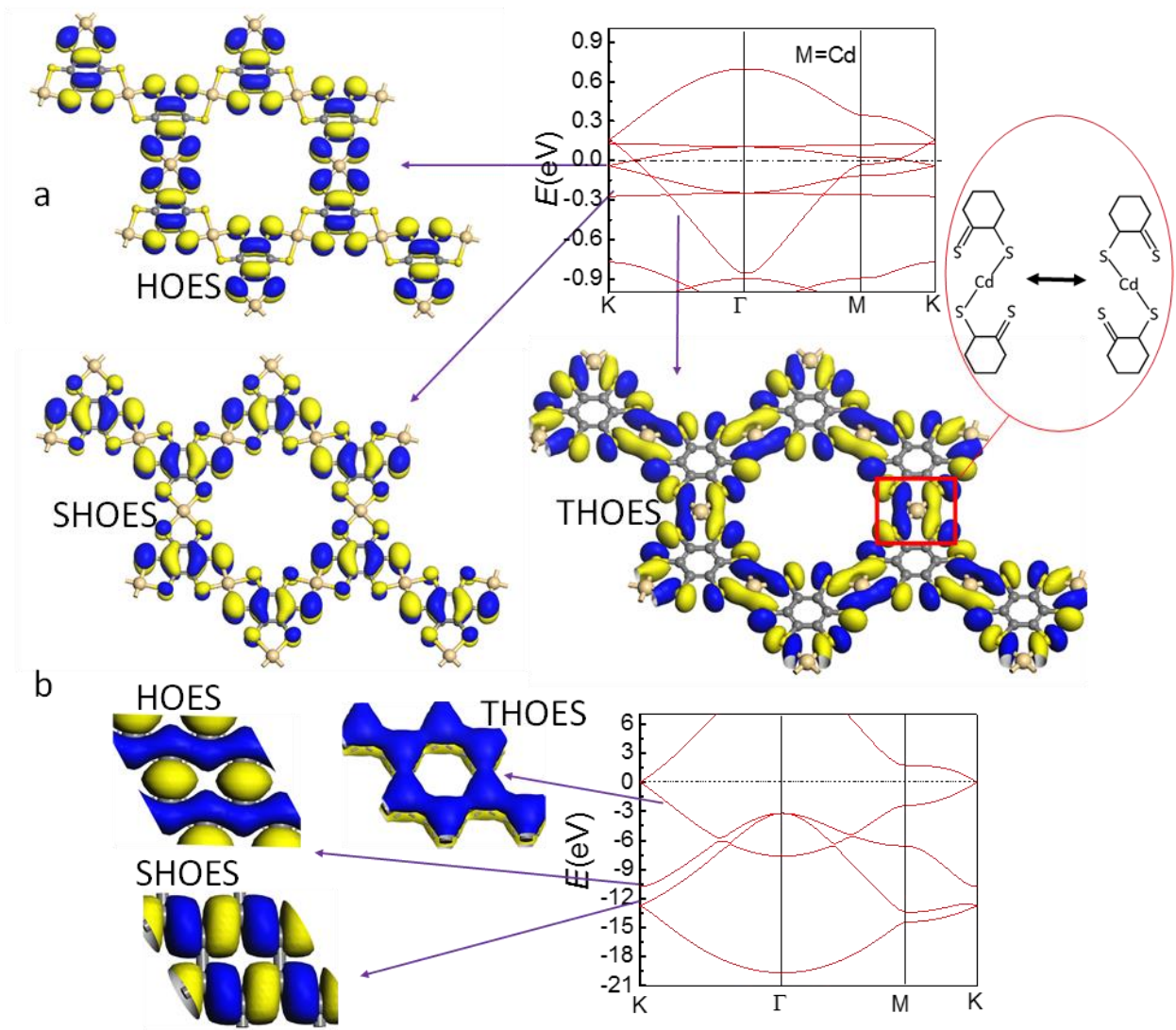

Analysis of electron states of (a) $Cd_3C_{12}S_{12}$ compared with (b) graphene.

We have investigated the origin of zero hybridization between the fast and slow Dirac cones. As a prototype, in Fig.2a we plot the highest occupied electron state (HOES), second and third highest occupied electron state (SHOES and THOES) at Γ point for $Cd_3C_{12}S_{12}$ in comparison with graphene (Fig.2b). Surprisingly, we find that the fast Dirac cone (THOES) has contribution only from the $p_x$ and $p_y$ orbitals of sulfur, which is scarcely distributed on C or Cd atoms. This electronic state is also delocalized with high symmetry resembling the π electron of graphene (THOES), Here in the big red circle, the two configurations are equivalent and should have equal energy level. The THOES is highly delocalized between adjacent S atoms, where electron delocalization moves THOES away from Cd atoms, which may indicate the resonance between two configurations in the red circle.

The slow cone (HOES) is made up primarily from the $p_z$ orbitals of S and C, similar to SHOES (σ band) of graphene. Judging from the symmetry of electron states for the fast and slow Dirac cones, either at the Gamma point or at the crossing point of cones along Γ–K (Supplementary Fig.S2), their Dirac crossings are symmetry-protected in the absence of SOC. According to the symmetry of SHOES (flat bands), their crossing with both fast and slow Dirac cones can also be symmetry-protected, which can be observed in the band structure. Therefore those flat bands, slow cones, fast cones are completely independent channels and will not mix with each other. The wavefunction of slow cones (HOES) corresponds to the configuration where C=C double bonds are formed favoring a smaller bond length compared with C-C bonds. This explains why upon biaxial tensile strain the slow cones will rise in energy relative to fast cones in Fig.1.

It is intriguing that all the bands near the Fermi level contain no contribution from the metal orbitals, but from C or S atoms only, so SOC near the Fermi energy induced by these light elements is negligibly small. Indeed, even for the heaviest M = Hg, the bandgap induced by SOC is within 2.9 meV. It is especially intriguing that the orbital of the fast Dirac cone finds contribution only from the S atoms, where every four S atoms (as marked in red rectangle) share two s electron donated from Cd, and the delocalized orbital between two nearest S atoms indicates the formation of multicenter electron-deficient covalent bonds. The non-hybridization between the fast and slow cones is due to the symmetry of the 2D atomic sheets and the orthogonality between the $p_x$ - $p_y$ and $p_z$ orbital manifolds.

**Figure 3 Bandstructures of $M_3C_{12}O_{12}$**

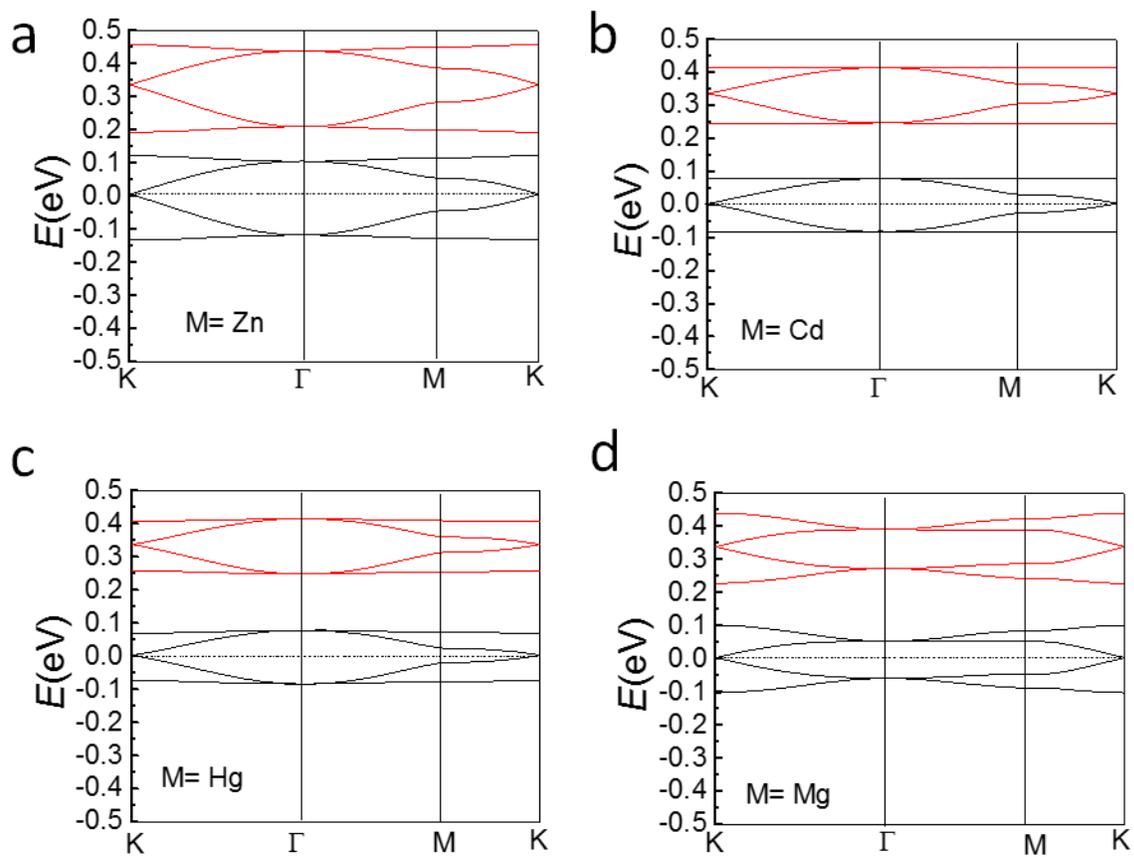

(a-d) Band structures of $M_3C_{12}O_{12}$ for M= Zn, Cd, Hg, and Mg. Black and red lines represent spin-up and spin-down channel, respectively.

**Figure 4 Other 2D MOFs with different metal atoms or molecules.**

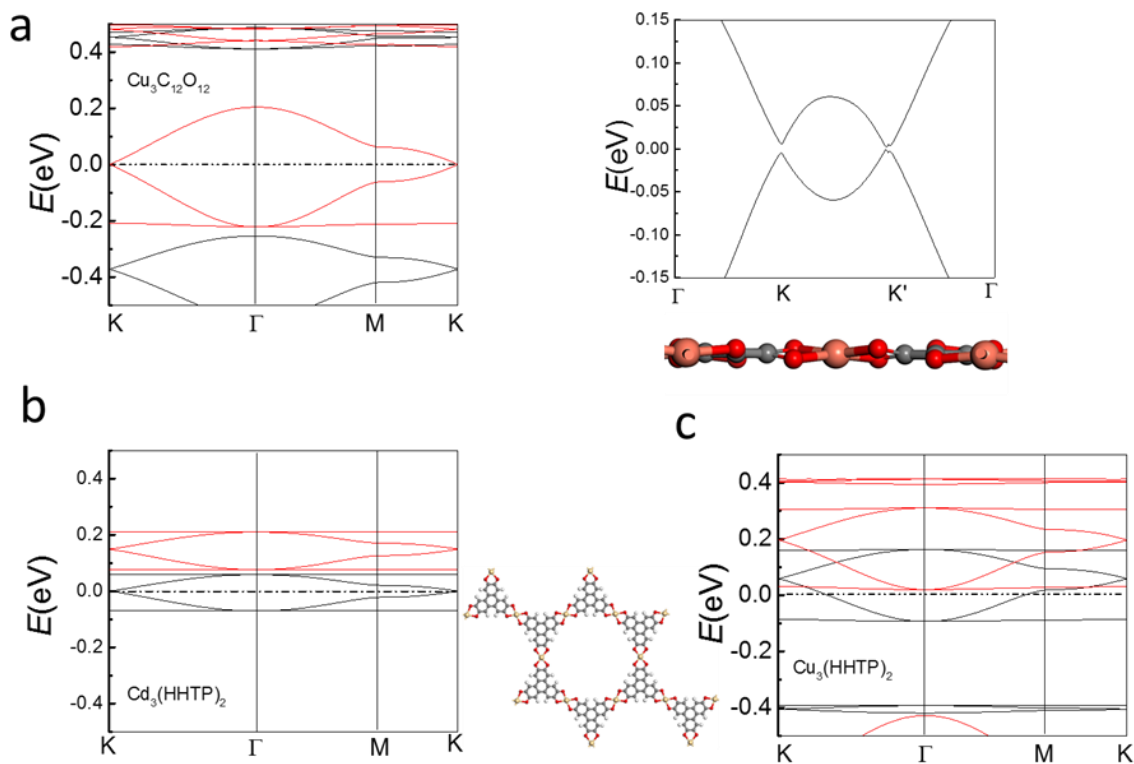

Band structures of (a) $Cu_3C_{12}O_{12}$, with and without SOC, (b) $Cd_3(HHTP)_2$ and (c) $Cu_3(HHTP)_2$.

It is evident from the literature that when $Zn^{2+}$, $Cd^{2+}$, and $Hg^{2+}$ are coordinated by two 1,2-benzenedithiolates molecules [26], the resulting coordination complexes tend to adopt tetrahedral geometry. The tetrahedral coordination environment maximizes the distance between coordination ligands, thus minimizing the electro-repulsive interaction between them. However, the square planar coordination environment of $Zn^{2+}$, $Cd^{2+}$, and $Hg^{2+}$ may also be possible if other stabilization energy compensates the electro-repulsion energy. For example, in the case of $Cd_3C_{12}S_{12}$, the S-S bonding interaction in a [$CdS_4$] unit and the in-plane conjugation throughout the 2D sheet (Fig. 2a THEOS) both stabilize the square planar coordination environment of $Cd^{2+}$. As a result, the $Cd_3C_{12}S_{12}$ sheet is likely to be planar, especially under tensile strain. A similar stabilization effect could also lead to planar ground states of the other 2D MOFs proposed here.

**Magnetic Half Dirac cones and QAHE in $M_3C_{12}O_{12}$ and other 2D MOFs**

Replacing sulfur by oxygen is chemically feasible and was investigated here computationally as well. Remarkably, we find that $M_3C_{12}O_{12}$ exhibit dissimilar band structures compared with $M_3C_{12}S_{12}$. As shown in Fig.3(a-d), $M_3C_{12}O_{12}$ (M = Zn, Cd, Hg, Mg) all become spontaneously spin-polarized with a magnetic moment of $2\mu_B$ per supercell, which is evenly distributed on C and O atoms but scarcely on M. That is, they are ferromagnetic monolayers in their native undoped condition. According to the Stoner criterion $\rho(E_F)U \geq 1$, where $U$ is the Hubbard correlation energy, spontaneous spin-polarization (ferromagnetism) can be induced by the high density of states at the Fermi level $\rho(E_F)$. In spin-restricted calculations, the flat bands are located right near the Fermi level in $Cd_3C_{12}O_{12}$, for example (Supplementary Fig. S4a), leading to a very large $\rho(E_F)$ that can give rise to ferromagnetism. Similar mechanism may also apply on the induced ferromagnetism of $Mg_3C_{12}S_{12}$ upon strain as mentioned above, where the bands become more localized in Fig. S1. We can estimate the Curie temperature $T_c$ by using mean-field theory of Heisenberg model, $T_c = \frac{2\Delta}{3k_B}$, where $\Delta$ is the energy cost to flip one spin in ferromagnetic lattice. Here we can regard every $C_6O_6$ unit as a single spin, and for the case of $Zn_3C_{12}O_{12}$, $\Delta$=27meV and $T_c$=209K. Due to the spin-splitting, the slow Dirac cones in one spin channel (half Dirac cones) place the Dirac points right at the Fermi level, while the slow Dirac cones in the other spin channel together with the fast Dirac cones are pushed away from the Fermi level. As a result, those systems are conducting in only one spin channel and one cone channel. Taking M = Cd as an example, the plot of HOES and SHOES of $Cd_3C_{12}O_{12}$ in Fig. S3 have a perfect match with HOES and SHOES of $Cd_3C_{12}O_{12}$ in Fig. 2 corresponding to their slow Dirac cones, which are also distributed by O and C atoms, so SOC induced bandgap is even more negligible (Supplementary Table S1).

Half Dirac cones may still emerge upon substitution of M with Cu, although the structure will become buckled just like $Cu_3(HITP)_2$ [21], which has been synthesized recently [27]. As shown in Fig. 4a, $Cu_3C_{12}O_{12}$ exhibits similar half Dirac cones, with $1\mu_B$ magnetic moment per supercell. If SOC is taken into account, due to the breaking of inversion symmetry induced by buckling, a bandgap as large as 5.3 meV at K' and 10.2 meV at K will open up in the band structure of $Cu_3C_{12}O_{12}$ and will give rise to 2D quantum anomalous Hall effect (QAHE)[28]. Actually, all the systems with half Dirac cones in this paper will exhibit QAHE if the SOC strength is not negligible. To verify this, we calculate the Chern number by a brief formula since they all have $C_3$ rotational symmetry:

$$e^{-i\pi C/3} = \prod_{i=1}^{n} e^{i\pi} \lambda_i(\Gamma)\lambda_i(K)\lambda_i(K')$$

where $\lambda_i$ is the $C_3$ eigenvalues of the *i*-th band and *n* is the number of the valence bands. All the eigenvalues of the occupied states with marked band index have been obtained by first-principles calculations, as shown in Supplementary Fig. S5. The Chern number turns out to be -1 and 1 for $Cu_3C_{12}O_{12}$ and $Cd_3C_{12}O_{12}$, respectively, which signifies the existence of QAHE state. To further enhance the chance of experimental realization of our predictions, we investigate more similar 2D MOFs that have been fabricated recently, for example those synthesized from 2,3,6,7,10,11-hexahydroxytriphenylene ($H_{12}C_{18}O_6$, HHTP)[16] and various metal ions, which assemble into 2D porous extended frameworks, which may exhibit similar nontrivial properties. As shown in Fig. 4c, $Cd_3(HHTP)_2$ exhibits a half Dirac cone at each K point similar to $M_3C_{12}O_{12}$ in Fig .3. For $Cu_3(HHTP)_2$, which has been fabricated in Ref. 15, although the Dirac points do not coincide with the Fermi level, we find that it is still conducting in only one cone channel and only one spin channel.

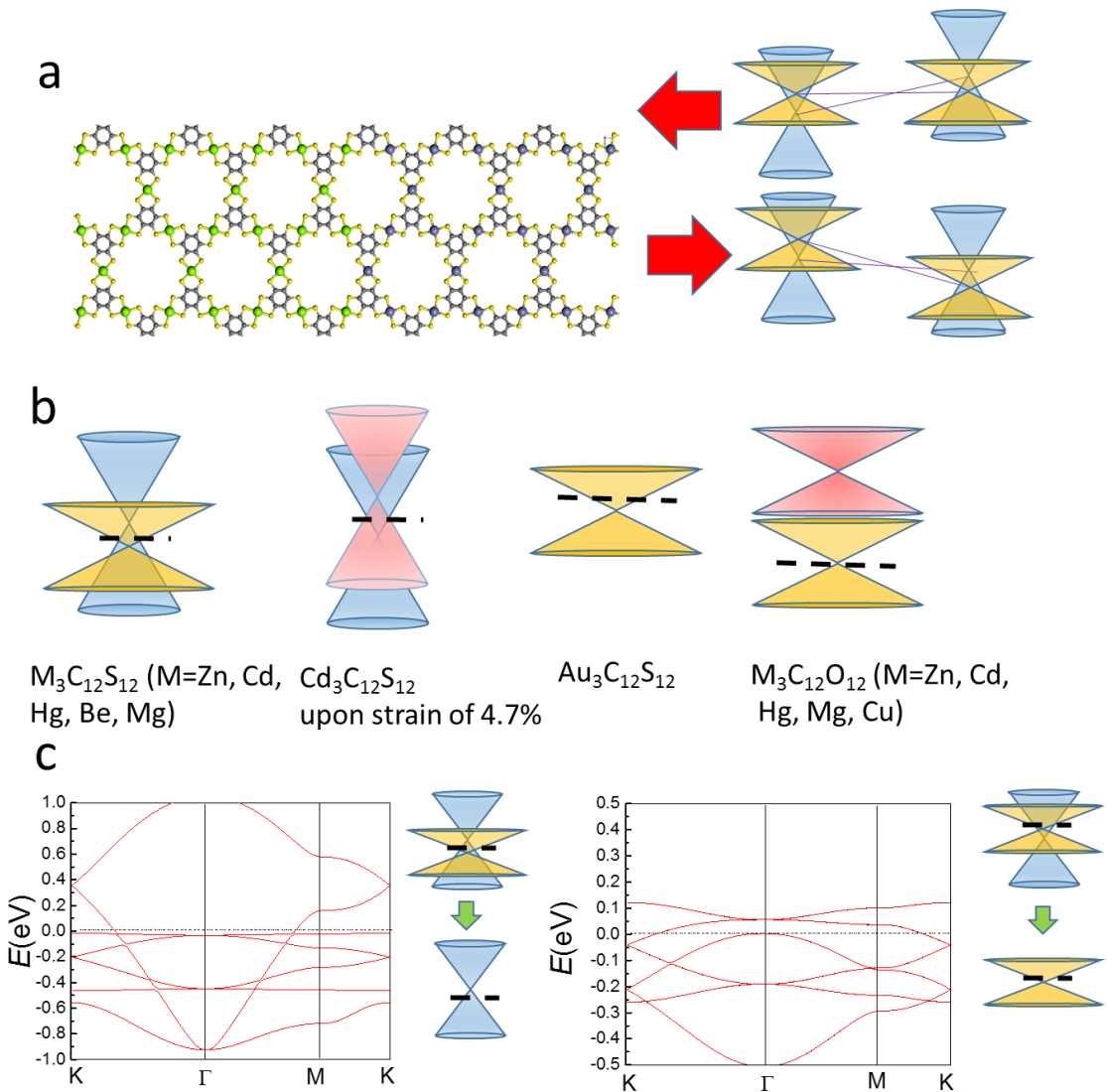

**Figure 5 Heterostructures and conetronics devices**

(a) Heterostructures composed of different 2D MOFs. (b) Cone selecting components. Band structures of (c) $Hg_3C_{12}S_{12}$ and (d) $Mg_3C_{12}S_{12}$ upon charge injection of 2e and 2.4e per supercell (containing 27atoms) respectively.

**Design of heterostructures and cone selecting devices**

The lattice constants of $M_3C_{12}S_{12}$ are approximately the same for M = Cd, Hg, Mg (see Table S1, variations are all within 0.5%), and also for M = Zn, Ni, Be (around 2.0% difference, in which $Ni_3C_{12}S_{12}$ is semiconductor with a Dirac cone 0.5eV above the undoped Fermi level[15]). This good match in lattice constants may give rise to the construction of a variety of heterostructures with different Dirac fermions, motivated by "the interface is the device"[29, 30]. For example, as shown in Fig. 5a, a p-n Dirac fermion junction can be constructed from $Mg_3C_{12}S_{12}$ with p-type fast cone / n-type slow cone and $Cd_3C_{12}S_{12}$ (or $Hg_3C_{12}S_{12}$) with n-type fast cone / p-type slow cone. Upon an electric forward bias to the left, the slow cones will be pushed towards the same energy level for both sides while the fast cones are pushed away, so the favorable conducting channel will be the slow cones; upon an electric forward bias to the right, the fast cones will be pushed towards the same energy level for both sides while the slow cones are pushed apart, so the favorable conducting channel will be the fast cones. This device can be used to switch the conducting channel between fast and slow cones, as illustrated in the graph. For $M_3C_{12}O_{12}$ material with slightly smaller lattice constant, $Hg_3C_{12}O_{12}$ can match and join in the heterostructure composed by $Zn_3C_{12}S_{12}$, $Be_3C_{12}S_{12}$, or $Ni_3C_{12}S_{12}$ (see Table S1, variations are all within 4%), which will bring in spintronic effect simultaneously. The half-metallic $Zn_3C_{12}O_{12}$ can also be connected to nonmagnetic metal $Ni_3C_{12}O_{12}$ (see Fig. S4, 3% difference in lattice constant) for spin-injection. In fact, for any of these devices, the heterostructures can be either in-plane or vertical, referring to previous design of all-metallic vertical transistors based on stacked Dirac materials, with Dirac cones in different positions of k-space due to bilayer twist or in-plane strains [31].

We note that the fast and slow cones represent two independent channels for Dirac fermions at the same k-point, resembling spin-channels in spintronics, but with a new kind of pseudospin ($p_x$ - $p_y$ π-bonding versus $p_z$ π-bonding) instead of true spin. Such "conetronics" device can be realized if we can design components for selecting carriers of different cones. Our results have already shown that $M_3C_{12}S_{12}$ is conducting in both cone channels (and in both spin channels) – see Fig. 1 – but that $M_3C_{12}O_{12}$ is metallic only in the slow cone channel (and only for the spin-up channel) – see Fig. 3. The next desirable component should be only conducting in the fast cone channel, and we find that $Cd_3C_{12}S_{12}$ indeed meets this requirement upon a biaxial strain of 4.7%. This strain is still well within the realm of experimental feasibility: the spin-splitting in the band structure (see Fig. S4) pushes the slow cones away from the Fermi level leaving only the fast cones conducting. However, upon this strain the lattice constant of $Cd_3C_{12}S_{12}$ (16.29 Å) will be much larger than that of $M_3C_{12}O_{12}$. The band structure of $Au_3C_{12}S_{12}$ calculated in Ref. 16 is conducting only in the slow cone channel, and maintains this property when stretched to the same lattice constant (see Fig. S4). These conetronic components are summarized in Fig.5b.

An alternative way for conetronic manipulation is electrostatic charge injection, as electrostatic gate doping is highly feasible in 2D bulk monolayers but not so in 3D bulk materials due to Coulomb explosion. As shown in Fig. 5c, $Hg_3C_{12}S_{12}$ will become metallic only in the fast cone channel upon a charge injection of $2e^-$ per supercell, while $Mg_3C_{12}S_{12}$ will be conducting only in the slow cone channel upon a charge injection of $2.4e^-$ per supercell. According to our calculations, electrostatic gate doping would allow one to manipulate the ferromagnetic properties of $M_3C_{12}O_{12}$ and $M_3(HHTP)_2$ monolayers, giving rise to electrical control of magnetic properties, which is a form of the long-sought-after "multi-ferroic" effect. For example, upon hole doping of $+2e^-$ per supercell, $Cd_3C_{12}O_{12}$ (27 atoms per supercell) and $Cd_3(HHTP)_2$ (63 atoms per supercell) will all become completely nonmagnetic semiconductors. Here it is worth to mention that Dirac field-effect transistors with ultra-high on/off ratio can be obtained: for example, upon a gate voltage, when the Fermi level of $Mg_3C_{12}S_{12}$ is raised by 0.3eV, or the Fermi level of $Cd_3C_{12}O_{12}$ and $Cd_3(HHTP)_2$ declines by just 0.1eV, they will switch from the relativistic high-mobility metallic state to the high-resistivity semiconducting state.

**Discussions**

Dilute magnetic semiconductors have been researched, with the aim that magnetic devices can be directly integrated in the current semiconductor-based circuits. Now we have proposed 2D MOFs with different properties: magnetic, nonmagnetic, semiconducting, metallic, fast cone channel conducting only, slow cone channel conducting only, both cone channels conducting, etc. They are 2D MOFs with different M and may be directly integrated into a monolayer 2D MOF sheet, holding the promise to build multifunctional devices in the 2D MOF-based circuits.

In MOF based devices, fast Dirac fermions and slow Dirac fermions may be distinguished by their difference in carrier mobility, Landau energy levels, magneto-oscillation frequency, gate voltage-dependent effective mass, Klein tunneling (with different "speed of light"), referring to reports by Kim's group[32] and Geim's group[33, 34], where comparing their mobility should be the most convenient among them. An alternative way is to use "coneresistance" (CR), just like giant magnetoresistance (GMR) between spin-parallel state and anti-parallel state. For example, in Fig. 5c $Hg_3C_{12}S_{12}$ and $Mg_3C_{12}S_{12}$ can be regarded as in anti-parallel state upon charge injection, so when they constitute the heterojunction shown in Fig.5a, the charged state will be a high resistance state compared with neutral state. In Fig. S6 we compute their transmission spectrum, and it turns out that the charged state has a much lower transmission compared with neutral state. Similar to GMR, here CR=$\Delta R/R$=718%. Like the magnetic and electric degrees of freedom in multiferroics, we have already obtained spin and cone-pseudospin degrees of freedom in a number of 2D MOFs, that can be manipulated by strain, electric and magnetic fields.

Therefore, compared with spintronics and valleytronics, conetronics may have some distinct advantages. Besides non-saturating magnetoresistance, high mobility and low power consumption by Dirac Fermion carriers, its manipulation using electrical gate is more preferable compared with magnetic field or optical pumping. Besides, higher density data storage requires smaller sizes or lower-dimensions, where the problem of superparamagnetism or low Curie-temperature can always arise in spintronics. Here the applications could be at room temperature as the energy differences between the cones are larger than 100meV.

In summary, based on first-principles calculations, we have proposed a concept of conetronics in $M_3C_{12}S_{12}$ and $M_3C_{12}O_{12}$ where M = Zn, Cd, Hg, Be or Mg donating only two s electrons with no contribution to bands near Fermi level. For $M_3C_{12}S_{12}$, their band structures exhibit double Dirac cones with different velocity that are respectively n and p type, which are tunable for switching via strain. The fast and slow cones can represent two independent Fermion channels since the Dirac crossing of two cones are symmetry-protected, making the systems 2D node-line semimetals. The node line rings right at their crossing, which are both electron and hole pockets at the Fermi level, can give rise to magnetoresistance that will not saturate when magnetic field is infinitely large. While those $M_3C_{12}S_{12}$ are conducting in both cone channels and both spin channels, for $M_3C_{12}O_{12}$ the spin-polarized slow Dirac cones locate Dirac cones precisely at the Fermi level, making the systems conducting in only one spin channel and one cone channel. Some MOFs with considerable SOC like $Cu_3C_{12}O_{12}$ can even exhibit QAHE, while those based on metal-hexahydroxytriphenylenes can exhibit similar nontrivial properties. Finally, we design a variety of heterostructure devices composed of different $M_3C_{12}S_{12}$ and $M_3C_{12}O_{12}$ layers, giving rise to a variety of spintronic conetronics devices. We anticipate that electrostatic gating would allow one to manipulate the ferromagnetic properties of $M_3C_{12}O_{12}$ and $M_3(HHTP)_2$ monolayers, giving rise to electrical control of magnetic properties.

**Computational Methods**

First-principles calculations are performed within the framework of spin-unrestricted density-functional-theory (DFT) calculations implemented in the Vienna *ab initio* Simulation Package (VASP)[35]. The projector augmented wave (PAW) potentials[36] for the core and the generalized gradient approximation (GGA) in the Perdew-Burke-Ernzerhof (PBE)[37] form for the exchange-correlation functional are used. The Monkhorst-Pack *k*-meshes are set to 5 × 5 × 1 in the Brillouin zone and the nearest distance between two adjacent layers is set to 12 Å. All atoms are relaxed in each optimization cycle until atomic forces on each atom are smaller than 0.01 eV Å$^{-1}$ and the energy variation between subsequent iterations fall below 10$^{-6}$ eV. We also examine thermodynamic stability of $M_3C_{12}S_{12}$ and $M_3C_{12}O_{12}$ (take M=Cd for example) at 700 K using the Born-Oppenheimer molecular dynamics (BOMD) simulation[38], and the snapshots in Fig. S7 suggest that both systems will be stable. Transmission spectra are calculated by using the ATOMISTIX TOOLKIT (ATK) using non-equilibrium Green's function approach based the Landauer-Buttiker formula[39].

**Acknowledgements**

We acknowledge support by the Center for Excitonics, an Energy Frontier Research Center funded by the U.S. Department of Energy, Office of Science and Office of Basic Energy Sciences, under award number DE-SC0001088 (MIT). M.Wu and Z. Wang were supported by the National Natural Science Foundation of China (No. 21573084 and No. 11504117) and a start-up fund from Huazhong University of Science and Technology. H.H.F. is supported by the National Natural Science Foundation of China (No. 11274128). We thank Shanghai Supercomputer Center for making some of the computations possible. We are grateful for helpful discussions with Pablo Jarillo-Herrero, Marc Baldo, Wenbin Li, Xi Dai, Zengwei Zhu, Yingshuang Fu, Jingtao Lv and Jinghua Gao.

The authors declare no competing financial interest.

**Corresponding Author**

*Email: (M.H.W.)wmh1987@hust.edu.cn, (J.L.)liju@mit.edu

**Supporting Information**

Fermi velocity, lattice constant, bandstructure, orbital and rotational analysis, BOMD simulations of 2D MOFs, Figure S1-S7 and Table S1.